\documentclass{nature}
\usepackage{graphicx}
\usepackage{times}

\usepackage[utf8]{inputenc} 
\usepackage[T1]{fontenc}    
\usepackage{hyperref}       
\usepackage{url}            
\usepackage{booktabs}       
\usepackage{amsfonts}       
\usepackage{nicefrac}       
\usepackage{microtype}      
\usepackage{xcolor}
\usepackage{amsmath}
\usepackage{graphicx}
\usepackage{wrapfig}
\usepackage[ruled,vlined]{algorithm2e}
\SetKwComment{Comment}{$\triangleright$\ }{}
\SetCommentSty{itshape}
\usepackage{amsthm}
\usepackage{subfig}
\usepackage{array,multirow}
\usepackage{wrapfig}
\usepackage{changepage}

\usepackage{fancyhdr}
\title{Unsupervised deep learning identifies \\ semantic disentanglement \\ in single inferotemporal neurons}

\author
{Irina Higgins$^{1\dagger\ast}$, 
Le Chang$^{2,3\dagger}$, Victoria Langston$^{1}$,
Demis Hassabis$^{1,4}$\\
Christopher Summerfield$^{1,5}$\textsuperscript{\textdaggerdbl},  
Doris Tsao$^{2,6}$\textsuperscript{\textdaggerdbl}, Matthew Botvinick$^{1,4}$\textsuperscript{\textdaggerdbl}\\
\\
\normalsize{$^{1}$DeepMind, London, UK,}
\normalsize{$^{2}$Caltech, Pasadena, USA}\\
\normalsize{$^{3}$Chinese Academy of Sciences, Shanghai, China}\\
\normalsize{$^{4}$University College London, London, UK},
\normalsize{$^{5}$University of Oxford, Oxford, UK}\\
\normalsize{$^{6}$Howard Hughes Medical Institute, Pasadena, USA}\\
\\
\normalsize{$^\ast$To whom correspondence should be addressed; E-mail:  irinah@google.com}
\\
\normalsize{$^\dagger$Equal contribution, \textsuperscript{\textdaggerdbl}Equal contribution}
}

\begin{document} 


\baselineskip15pt


\maketitle 

\vspace{20pt}

\begin{abstract}
Deep supervised neural networks trained to classify objects have emerged as popular models of computation in the primate ventral stream. These models represent information with a high-dimensional distributed population code, implying that inferotemporal (IT) responses are also too complex to interpret at the single-neuron level. We challenge this view by modelling neural responses to faces in the macaque IT with a deep unsupervised generative model, \ensuremath{\beta}-VAE. Unlike deep classifiers, \ensuremath{\beta}-VAE ``disentangles'' sensory data into interpretable latent factors, such as gender or hair length. We found a remarkable correspondence between the generative factors discovered by the model and those coded by single IT neurons. Moreover, we were able to reconstruct face images using the signals from just a handful of cells. This suggests that the ventral visual stream may be optimising the disentangling objective, producing a neural code that is low-dimensional and semantically interpretable at the single-unit level. 
\end{abstract}

\section*{Introduction}
\paragraph{In search of a basic unit of representation in the neocortex.}
What is the basic unit of representation in the neocortex, and what computational objective gives rise to it? The foundational ``neuron doctrine'' argues that single cells are the key building blocks of brain function\cite{barlow1972neurondoctrine}, and decades of extracellular single neuron recordings have defined canonical coding principles, such as the sensitivity of early visual neurons to oriented contours and more anterior ventral stream neurons to complex objects and faces\cite{Hubel_Wiesel_1959,Chang_Tsao_2017}. More recently however, advances in recording methods have permitted the simultaneous recording of large populations of neurons\cite{saxena2019towards}. Hand-in-hand with this innovation has come the idea that meaningful variables (e.g. the gender of face) are encoded not in single neurons but in neural populations\cite{saxena2019towards,eichenbaum2018barlow,yuste2015neuron}, and that previously reported one-to-one mappings between the declarative aspects of the external world and single neurons may be spurious or misleading\cite{yuste2015neuron}. 

In parallel, visual neuroscience moved beyond handcrafted computational models towards theories that emphasise representation learning through end-to-end optimization\cite{richards2019framework,Yamins_Dicarlo_2016}. When trained with high-density teaching signals, contemporary deep networks can outperform humans on multiway object recognition tasks\cite{he2015imagenet}, and in doing so form high-dimensional representations that are multiplexed over many simulated neurons. Examined at the population level, these tuning distributions closely resemble those in biological systems\cite{khalighrazavi2014deep}, especially in higher-performing networks\cite{Yamins_etal_2014}, allowing deep learning networks to make accurate predictions about neural responses to synthesised images\cite{bashivan2018neural}. A natural synergy has thus arisen between new tools for multivariate population encoding and new computational theories that assume that the tuning properties of a single unit are all but uninterpretable\cite{richards2019framework,yuste2015neuron,eichenbaum2018barlow}. 

\paragraph{Disentangled representation learning through self-supervision.}
An important challenge for theories that rely on deep supervised networks, however, is that external teaching signals are scarce in the natural world, and visual development relies heavily on untutored statistical learning\cite{slone2015infants,lindsay_2020,thompson2016howcan}. Building on this intuition, one longstanding hypothesis\cite{Bengio_etal_2013,dicarlo2012how} is that the visual system uses self-supervision to recover the semantically interpretable latent structure of sensory signals, such as the shape or size of an object, or the gender or age of a face image. While appearing deceptively simple and intuitive to humans, such interpretable structure has proven hard to recover in practice, since it forms a highly complex non-linear transformation of pixel-level inputs. Recent advances in machine learning, however, have offered an implementational blueprint for this theory with the advent of deep self-supervised generative models that learn to ``disentangle'' high-dimensional sensory signals into meaningful factors of variation. One such model, known as the beta-variational autoencoder (\ensuremath{\beta}-VAE), learns to faithfully reconstruct sensory data from a low-dimensional embedding whilst being additionally regularised in a way that encourages individual network units to code for semantically meaningful variables, such as the colour of an object, the gender of a face, or the arrangement of a scene (Fig. 1a-c)\cite{Higgins_etal_2017,burgess2019monet,lee2020idgan}. These deep generative models thus continue the longstanding tradition from the neuroscience community of building self-supervised models of vision\cite{fukushima1980neocognitron,riesenhuber1999hmax}, while moving in a new direction that allows strong generalisation, imagination, abstract reasoning, compositional inference and other hallmarks of biological visual cognition\cite{Higgins_etal_2017b,Higgins_etal_2018,burgess2019monet,Achille_etal_2018}.

\section*{Results}

\paragraph{How well do single disentangled latent units explain the responses of single neurons?}
If the computations employed in biological sensory systems resemble those employed by this class of deep generative model to disentangle the visual world, then contrary to the ``population doctrine''\cite{saxena2019towards,eichenbaum2018barlow,yuste2015neuron}, the tuning properties of single neurons should map readily onto the meaningful latent units discovered by the \ensuremath{\beta}-VAE. Here, we tested this hypothesis, drawing on a previously published dataset\cite{Chang_Tsao_2019} of neural recordings from 159 neurons in macaque face area AM, made whilst the animals viewed 2,100 natural face images (Fig. 2a, see Online Methods). Using face perception as the test domain for understanding whether IT neurons may be employing similar disentangling learning mechanisms to the deep generative models has unique advantages. Specifically, both neural responses and image statistics in this domain have been particularly well studied compared to other visual stimulus classes. This allows for comparisons with strong hand-engineered baselines\cite{Chang_Tsao_2017} using relatively densely sampled neural data\cite{tsao2008mechanisms}. Furthermore, although faces make up a small subset of all possible visual objects, and neurons that preferentially respond to faces tend to cluster in particular patches of the inferotemporal (IT) cortex\cite{tsao2008mechanisms}, the computational mechanisms and basic units of representation employed for face processing may in fact generalise more broadly within the ventral visual stream\cite{tarr2000ffa,tsao2008mechanisms}.

We first investigated whether the variation in average spike rates of any of the individual recorded neurons was strongly explained by the activity in single units of a trained \ensuremath{\beta}-VAE that learnt to ``disentangle'' the same face dataset that was presented to the primates. For illustration, in Fig. 1c we show faces that were generated (or ``imagined'') by such a \ensuremath{\beta}-VAE. Each row of faces is produced by gradually varying the output of a single network unit (we call these ``latent units''), and it can be seen that they learnt to encode interpretable variables -- e.g. hairstyle, age, face shape or emotional variables such as the presence of a smile. Individual disentangled units discovered by the \ensuremath{\beta}-VAE were also able to explain the response variance in single recorded neurons, as shown in Fig. 2b. For example, neuron 95 is shown to be sensitive to the thickness of the hair, and neuron 136 is shown to respond differentially to the presence of a smile.

To quantify this effect, we used a metric recently proposed in the machine learning literature, referred to as neural ``alignment''\footnote{Two versions of the same measure were simultaneously and independently proposed in the machine learning literature, referred to as ``completeness''\cite{Eastwood_Williams_2018} or ``compactness''\cite{Ridgeway_Mozer_2018}. We choose to refer to the same measure as ``alignment'' for more intuitive exposition.} in this work, which measures the extent to which variance in each neuron's firing rate can be explained by a single latent unit\cite{Eastwood_Williams_2018}, but is insensitive to the converse, i.e. whether a single unit predicts the response of many biological neurons (Fig. 3a, see Online Methods). High alignment scores thus indicate that a neural population is intrinsically low-dimensional, with the factors of variation mapping onto the variables discovered by the latent units of the neural network. We first compared alignment scores between the \ensuremath{\beta}-VAE and the monkey data to a theoretical ceiling which was obtained by subsampling the neural data to match the intrinsic dimensionality of the \ensuremath{\beta}-VAE latent representation (see Online Methods) and computing its alignment with itself (Fig. 3b). Remarkably, alignment scores in the \ensuremath{\beta}-VAE met this ceiling, with no reliable difference between the two estimates obtained when the analysis was repeated on multiple subsamples and with multiple network instances ($p = 0.43$, Welch’s t-test). Furthermore, when we repeated this analysis while computing alignment against fictitious neural responses obtained by linearly recombining the original neural data, we found a significant drop in scores for both the \ensuremath{\beta}-VAE and neural subsets (Fig. 3c, $p < 0.01$, Welch’s t-test), indicating that the individual disentangled units discovered by the \ensuremath{\beta}-VAE map significantly better onto the responses of single neurons recorded from macaque IT, rather than onto their linear combinations.

The extent to which the \ensuremath{\beta}-VAE is effective in disentangling a dataset into its latent factors can vary substantially with the way it is regularised, as well as with randomness in its initialisation and training conditions\cite{Locatello_etal_2019}. The  parameter after which the network class is named determines the weight of a regularisation term that aims to keep the latent factors independent. Networks with higher values of $\beta$ thus typically give rise to more disentangled representations, as measured by a metric known as the unsupervised disentanglement ranking (UDR, see Online Methods)\cite{Duan_etal_2019}, a finding we replicate here. However, we also found that networks with higher UDR scores additionally had higher alignment scores with the neural data (Fig. 3d), and that this relationship held for networks with the same and different values of $\beta$ (Fig. 3e).  In other words, the better the network was able to disentangle the latent factors in the face dataset, the more those factors were expressed in single neurons recorded from macaque IT.

\paragraph{No single aspect of the disentanglement objective is sufficient to achieve high alignment with neural responses}
Next, we compared the \ensuremath{\beta}-VAE alignment scores with a number of rival models. These baseline models were carefully chosen to disambiguate the role played by the different aspects of the \ensuremath{\beta}-VAE design and training in explaining the coding of neurally aligned variables in its single latent units (see Online Methods). We included a state-of-the-art deep supervised network (VGG,\cite{Parkhi_etal_2015}) that has previously been proposed as a good model for comparison against neural data in face recognition tasks\cite{grossman2019evolution,dobs2019face}, other generative models, such as a basic autoencoder (AE)\cite{Hinton_Salakhutdinov_2006} and a variational autoencoder (VAE)\cite{Kingma_Welling_2014}, as well as baselines provided by ICA, PCA and a classifier which used only the encoder from the \ensuremath{\beta}-VAE. We defined ``latent units'' as those emerging in the deepest layers of these networks and, where appropriate, used PCA or feature subsampling (e.g. for VGG raw) to equate the dimensionality of the latent units (to $\leq 50$) to provide a fair comparison with the \ensuremath{\beta}-VAE. We also compared \ensuremath{\beta}-VAE to the ``gold standard'' provided by the previously published active appearance model (AAM)\cite{Chang_Tsao_2017}, which produced a low-dimensional code that explained the responses of single neurons to face images well\cite{Chang_Tsao_2017,Chang_Tsao_2019}. Unlike the \ensuremath{\beta}-VAE, which relied on a general learning mechanism to discover its latent units, AAM relied on a manual process idiosyncratic to the face domain. Hence, \ensuremath{\beta}-VAE provides a \emph{learning}-based counterpart to the handcrafted AAM units that could generalise beyond the domain of faces. Although the baselines considered varied in their average alignment scores (Fig. 4a), none approached those of the \ensuremath{\beta}-VAE, for which alignment was statistically higher than every other model (all p-values $< 0.01$, Welch’s t-test). The alignment scores broken down by individual neurons are plotted in Fig. 4b for the \ensuremath{\beta}-VAE and its baselines. 

We validated the findings above using a more direct metric for the coding of latent factors in single neurons, which compared the ratio between the maximum correlation between spike rates and activations in each latent unit, and the sum of such correlations over the model units (average correlation ratio in Fig. 5a, see Online Methods). This ratio was higher for the \ensuremath{\beta}-VAE than for other models, confirming the results with alignment scores (Fig. 5b). Interestingly, different neurons did not tend to covary with the same \ensuremath{\beta}-VAE latent unit. In fact, there was more heterogeneity among \ensuremath{\beta}-VAE units that achieved maximum correlation with the neural responses than among the equivalent units for other models (Fig. 5c).  Rich heterogeneity in response properties of single neurons (or latent units) is exactly what would be desired to enable a population of computational units to encode the rich variation in the image dataset. 

Taken together these results suggest that no one feature of the \ensuremath{\beta}-VAE  -- its architecture (baselined by AE, VAE and classifier), training data distribution (baselined by VGG) or isolated aspects of its learning objective (baselined by PCA and ICA) -- was sufficient to explain the coding of neurally aligned latent variables in single units. Rather, it was all of these design choices together that allowed the \ensuremath{\beta}-VAE to learn a set of disentangled latent units that explained the responses to single neurons so well.

\paragraph{Disentangled units carry sufficient information to decode previously unseen faces from as few as twelve neurons}
Finally, we conducted an analysis that sought to link the virtues of the \ensuremath{\beta}-VAE as a tool in machine learning -- its capacity to make strong inferences about held out data -- with its qualities emphasised here as a theory of visual cognition -- strong one-to-one alignment between individual neural and individual disentangled latent units. During training we omitted 62 faces that had been viewed by the monkeys from the training set of the \ensuremath{\beta}-VAE, allowing us to verify that these were reconstructed more faithfully by the \ensuremath{\beta}-VAE than by other networks. Critically, in order to reconstruct these faces, we applied the decoder of the \ensuremath{\beta}-VAE not to its latent units as inferred by its encoder, but rather to the latent unit responses predicted from the activity of a small subset of single neurons (as few as twelve) that best aligned with each model unit on a different subset of data (Fig. 6a, see Online Methods). We found that such one-to-one decoding of latent units from the corresponding single neurons was significantly more accurate for the disentangled latent units learnt by the \ensuremath{\beta}-VAE compared to the latent units learnt by other baseline models (all p-values $< 0.01$, Welch’s t-test) (Fig. 6b). Furthermore, we visualised the \ensuremath{\beta}-VAE reconstructions decoded from just twelve matching neurons (Fig. 6c). Qualitatively, these appeared both more identifiable and of higher image quality than those produced by the latent units decoded from the nearest rival models, the AE and the basic VAE, which required twice as many neurons for decoding (Fig. 6c). It should be noted that the AE was explicitly optimised for reconstruction quality, while the \ensuremath{\beta}-VAE was optimised for disentangling. These results suggest that both the small subset of just twelve neurons and the corresponding twelve disentangled units carried sufficient information to decode previously unseen faces, the capacity that is required for effective vision in an unpredictable and ever-changing natural world.

\section*{Discussion}
\paragraph{Disentangled representation learning as a predictive computational model of vision.}
The results we have presented here provide evidence that the code for facial identity in the primate IT may in fact be low-dimensional and interpretable at a single neuron level. In particular, we showed that the axes of variation represented by single IT neurons align with single semantically meaningful ``disentangled'' latent units discovered by the \ensuremath{\beta}-VAE, a recent class of self-supervised deep neural networks proposed in the machine learning community.

Our work extends recent studies of the coding properties of single neurons in the primate face patch area, reporting finding one-to-one correspondences between model units and neurons, as opposed to few-to-one as previously reported\cite{Chang_Tsao_2017}. Moreover, we show that disentangling may occur at the end of the ventral visual stream (IT), extending results recently reported for V1\cite{gaspar2019representational}. Past studies have proposed that the ventral visual cortex may disentangle\cite{dicarlo2012how,gaspar2019representational} and represent visual information with a low-dimensional code\cite{Op_de_Beeck_etal_2001,Kayaert_etal_2005,Chang_Tsao_2017}. However, this work did not ask how these representations emerge via learning. Here, we propose a theoretically grounded\cite{Higgins_etal_2018b} computational model (the \ensuremath{\beta}-VAE) for how disentangled, low-dimensional codes may be learnt from the statistics of visual inputs\cite{Higgins_etal_2017}. 

An important aspect of our proposed learning mechanism is that it generalises beyond the domain of faces\cite{Higgins_etal_2017,burgess2019monet,lee2020idgan}. We believe that the difficulty in identifying interpretable codes in the IT encountered in the past may have been due to the fact that semantically meaningful axes of variation of complex visual objects are more challenging for humans to define (and hence use as visual probes) compared to simple features, such as visual edges\cite{lindsay_2020}. A computational model like the \ensuremath{\beta}-VAE, on the other hand, is able to automatically discover disentangled latent units that align with such axes, as was demonstrated for the domain of faces in this work. Hence, assuming that the computational mechanisms underlying face perception in the brain generalise to the broader set visual domains\cite{tarr2000ffa,tsao2008mechanisms}, \ensuremath{\beta}-VAE may serve as a promising tool to understand IT codes at a single neuron level even for rich and complex visual stimuli in the future. 

One contribution of this paper is the introduction of novel measures for comparing neural and model representations. Unlike other often used representation comparison methods which are insensitive to invertible linear transformations\cite{Kriegeskorte2008rsa,Yamins_etal_2014}, our methods measure the alignment between individual neurons and model units. Hence, they do not abstract away the representational form and preserve the ability to discriminate between alternative computational models that may otherwise score similarly\cite{thompson2016howcan}.

While the development of \ensuremath{\beta}-VAE for learning disentangled representations was originally guided by high level neuroscience principles\cite{Wood_Wood_2018,Smith_etal_2018,Friston_2010}, subsequent work in demonstrating the utility of such representations for intelligent behaviour was primarily done in the machine learning community\cite{Higgins_etal_2017b,Higgins_etal_2018,Achille_etal_2018}. In line with the rich history of mutually beneficial interactions between neuroscience and machine learning\cite{Hassabis_etal_2017}, we hope that the latest insights from machine learning may now feed back to the neuroscience community to investigate the merit of disentangled representations for supporting intelligence in the biological systems, in particular as the basis for abstract reasoning\cite{Bellmund_etal_2018}, or generalisable and efficient task learning\cite{niv2019representations}.

\section*{Acknowledgments}
We would like to thank Raia Hadsell, Zeb Kurth-Nelson and Koray Kavukcouglu for comments on the manuscript. 


\newpage
\bibliography{scibib}

\newpage

\section*{Online methods}

\subsection*{Dataset}
We used a dataset of 2,162 natural grayscaled, centered and cropped images of frontal views of faces with neutral facial expressions pasted on a gray 200x200 pixel background as described in\cite{Chang_Tsao_2019}. The face images were collated from multiple publicly available datasets\cite{Martinez_Benavente_1998,Liu_etal_2015,Ma_etal_2015,Peer_1999,Phillips_etal_1998,Strohminger_etal_2016,Gao_etal_2008}. 62 held out face images were randomly chosen. These faces were among the 2,100 faces presented to the primates, but not among the 2,100 faces used to train the models. All models (apart from VGG) were trained on the same set of faces, which were mirror flipped with respect to the images presented to the primates. This ensured that the train and test data distributions were similar, but not identical. To train the Classifier baseline, we augmented the data with 5x5 pixel translations of each face to ensure that multiple instances were present for each unique face identity. The data was split into 80\%/10\%/10\% train/validation/test sets.

\subsection*{Neurophysiological data}
All neurophysiological data was re-used from\cite{Chang_Tsao_2019}. The data was collected from two male rhesus macaques (Macaca mulatta) of 7-10 years old. Face patches were determined by identifying regions responding significantly more to faces than to non-face stimuli while passively viewing images on a screen in a 3T TIM (Siemens, Munich, Germany) magnet. Tungsten electrodes (18–20~Mohm at 1~kHz, FHC) were used for single-unit recording. Spikes were sampled at 40~kHz. All spike data were re-sorted with offline spike sorting clustering algorithms (Plexon). Only well-isolated units were considered for further analysis. Monkeys were head fixed and passively viewed the screen in a dark room. Eye position was monitored using an infrared eye tracking system (ISCAN). Juice reward was delivered every 2–4~s if fixation was properly maintained. Images were presented in random order. All images were presented for 150~ms interleaved by 180~ms of a gray screen. Each image was presented 3–5 times. The number of spikes in a time window of 50-350~ms after stimulus onset was counted for each stimulus. See\cite{Chang_Tsao_2019} for further details.

\paragraph{Artificial neurophysiological data}
In order to investigate whether the responses of \ensuremath{\beta}-VAE units encoded linear combinations of neural responses, we created artificial neural data by linearly recombining the responses of the real neurons. We first standardised the responses of the 159 recorded neurons across the 2,100 face images. We then multiplied the original matrix of neural responses with a random projection matrix $A$. Each value $A_{ij}$ of the projection matrix was sampled from the unit Gaussian distribution. The absolute value of the matrix was then taken, and each column was normalised to sum to 1. 

\paragraph{Neuron subsets} For fairer comparison with the models, which learnt latent representations of size $N \in [10, 50]$ as will be described below, we sampled neural subsets with fifty or fewer neurons. To do this, we first uniformly sampled five values from $N \in [10, 50]$ without replacement to indicate the size of the subsets. Then, for each size value we sampled ten random neuron subsets without replacement, resulting in 50 neuron subsets in total.

\subsection*{Model details}

\subsubsection*{\ensuremath{\beta}-VAE model}
We used the standard architecture and optimisation parameters introduced in\cite{Higgins_etal_2017} for training the \ensuremath{\beta}-VAE (Fig. 7a). The encoder consisted of four convolutional layers (32x4x4 stride 2, 32x4x4 stride 2, 64x4x4 stride 2, and 64x4x4 stride 2), followed by a 256-d fully connected layer and a 50-d latent representation. The decoder architecture was the reverse of the encoder. We used ReLU activations throughout. The decoder parametrised a Bernoulli distribution. We used Adam optimiser with $1e-4$ learning rate and trained the models for 1~mln iterations using batch size of 16, which was enough to achieve convergence. The models were trained to optimise the following disentangling objective:

\begin{equation}
    \mathcal{L}_{\ensuremath{\beta}-VAE} = \mathbb{E}_{p(\mathbf{x})} [\ \mathbb{E}_{q_{\phi}(\mathbf{z}|\mathbf{x})} [\log\ p_{\theta}(\mathbf{x} | \mathbf{z})] - \beta KL(q_{\phi}(\mathbf{z}|\mathbf{x})\ ||\ p(\mathbf{z}))\ ]
\end{equation}

where $p(\mathbf{x})$ is the probability of the image data, $q(\mathbf{z}|\mathbf{x})$ is the learnt posterior over the latent units given the data, and $p(\mathbf{z})$ is the unit Gaussian prior with a diagonal covariance matrix.

\subsubsection*{Baseline models}
We compared \ensuremath{\beta}-VAE to a number of baselines to test whether any individual aspects of \ensuremath{\beta}-VAE training could account for the quality of its learnt latent units. To disambiguate the role of the learning objective, we compared \ensuremath{\beta}-VAE to a traditional \textbf{autoencoder (AE)}\cite{Hinton_Salakhutdinov_2006} and a basic \textbf{variational autoencoder (VAE)}\cite{Kingma_Welling_2014,Rezende_etal_2014}. These models had the same architecture, training data, and optimisation parameters as the \ensuremath{\beta}-VAE (Fig. 7a), but different learning objectives. The AE optimsed the following objective that tried to optimise the quality of its reconstructions:

\begin{equation}
    \mathcal{L}_{AE} = \mathbb{E}_{p(\mathbf{x})} \ || f(\mathbf{x}; \theta, \phi) - \mathbf{x} ||^2 
\end{equation}

where $f(\mathbf{x}; \theta, \phi)$ is the image reconstruction produced by putting the original image through the encoder and decoder networks parametrised by $\phi$ and $\theta$ respectively. The VAE optimised the variational lower bound on the data distrbution $p(\mathbf{x})$:

\begin{equation}
    \mathcal{L}_{VAE} = \mathbb{E}_{p(\mathbf{x})} [\ \mathbb{E}_{q_{\phi}(\mathbf{z}|\mathbf{x})} [\log\ p_{\theta}(\mathbf{x} | \mathbf{z})] - KL(q_{\phi}(\mathbf{z}|\mathbf{x})\ ||\ p(\mathbf{z}))\ ]
\end{equation}

where $q(\mathbf{z}|\mathbf{x})$ is the learnt posterior over the latent units given the data, and $p(\mathbf{z})$ is the isotropic unit Gaussian prior.

To test whether the supervised classification objective could be a good alternative to the self-supervised disentangling objective, we compared \ensuremath{\beta}-VAE to two classifier neural network baselines. One of these baselines, referred to as the \textbf{Classifier} in all figures and text, shared the encoder architecture, the data distribution and the optimisation parameters with the \ensuremath{\beta}-VAE (Fig. 5b), but instead of disentangling, it was trained to differentiate between the 2,100 faces using a supervised objective. In particular, the four convolutional layers and the fully connected layer of the encoder fed into an N-dimensional representation, which was followed by 2,100 logits that were trained to recognise the unique 2,100 face identities. In order to avoid overfitting, we used early stopping. The final models trained for between 300~k and 1~mln training steps. 

The other classifier baseline was the \textbf{VGG-Face} model\cite{Parkhi_etal_2015} (referred to as the VGG in all figures and text), a more powerful deep network developed for state-of-the-art face recognition performance and previously chosen as an appropriate computational model for comparison against neural data in face recognition tasks\cite{grossman2019evolution,Gucluturk2017reconstructing,dobs2019face} (Fig. 5c). Similarly to other works\cite{Chang_Tsao_2019,grossman2019evolution,Gucluturk2017reconstructing,dobs2019face}, we used a standard pre-trained VGG network, trained to differentiate between 2,622 unique individuals using a dataset of 982,803 images\cite{Parkhi_etal_2015}. Note that the data used for VGG training was unrelated to the 2,100 face images presented to the primates. The VGG therefore had a different architecture, training data distribution and optimisation parameters compared to the \ensuremath{\beta}-VAE. The model consisted of 16 convolutional layers, followed by 3 fully connected layers (see\cite{Parkhi_etal_2015} for more details). The last hidden layer before the classification logits contained 4,096 units. Following the precedent set by\cite{Chang_Tsao_2019} and\cite{Gucluturk2017reconstructing} we used PCA to reduce the dimensionality of the VGG representation by projecting the activations in its last hidden layer in response to the 2,100 test faces to the top N principal components (PCs) (Fig. 5c, referred to as VGG (PCA) in figures). Alternatively, we also randomly subsampled the units in the last hidden layer of VGG (without replacement) to control for any potential linear mixing of their responses which PCA could plausibly introduce (Fig. 7c, referred to as VGG (raw) in figures).

To rule out that the responses of single neurons could be modelled by simply explaining the variance in the data we compared \ensuremath{\beta}-VAE to N PCs produced by applying \textbf{principal component analysis (PCA)} to the 2,100 faces. To rule out the role of simply finding the independent components of the data during \ensuremath{\beta}-VAE training, we compared \ensuremath{\beta}-VAE to the N independent components discovered by \textbf{independent component analysis (ICA)} applied to the 2,100 face images. 

Finally, we also compared \ensuremath{\beta}-VAE to the \textbf{active appearance model (AAM)}. Linear combinations of small numbers of its latent units (six on average) was previously reported to explain the responses of single neurons in the primate AM area well\cite{Chang_Tsao_2017,Chang_Tsao_2019}. We re-used the AAM latent units from\cite{Chang_Tsao_2019}. These were obtained by setting 80 landmarks on each of the 2,100 facial images presented to the primates. The positions of landmarks were normalised to calculate the average shape template. Each face was warped to the average shape using spline interpolation. The warped image was normalised and reshaped to a 1-d vector. PCA was carried out on landmark positions and shape-free intensity independently. The first N/2 shape PCs and the first N/2 appearance PCs were concatenated to produce the N-dimensional AAM representations (Fig. 7d).

\subsubsection*{Training procedure and model selection}
To ensure that all models had a fair chance of learning a useful representation, we trained multiple instances of each model class using different hyperparameter settings. The choice of hyperparameters and their values were dependent on the model class. However, all models went through the same model selection pipeline: 1) $K$ model instances with different hyperparameter settings were obtained as appropriate; 2) $S \subseteq K$ models with the best performance on the training objective were selected; 3) models that did not discover any latent units that shared information with the neural responses were excluded, resulting in $M \subseteq S$ models retained for the final analyses. These steps are expanded below for each model class.

\paragraph{Hyperparameter sweep} For the \ensuremath{\beta}-VAE model the main hyperparameter of interest that affects the quality of the learnt latent units is the value of $\beta$. The $\beta$ hyperparameter controls the degree of disentangling achieved during training, as well as the intrinsic dimensionality of the learnt latent representation\cite{Higgins_etal_2017}. Typically a $\beta>1$ is necessary to achieve good disentangling, however the exact value differs for different datasets. Hence, we trained 400 models with different values of $\beta$ by uniformly sampling 40 values of $\beta$ in the $[0.5, 20]$ range. Another factor that affects the quality of disentangled representation is the random initialisation seed for training the models\cite{Locatello_etal_2019}. Hence, for each $\beta$ value, we trained 10 models from different random initialisation seeds, resulting in the total pool of 400 trained \ensuremath{\beta}-VAE model instances to choose from. All \ensuremath{\beta}-VAE models were initialised to have $N=50$ latent units, however due to the variability in the values of \ensuremath{\beta}, the intrinsic dimensionality of the trained models varied between ten and fifty. 

In order to isolate the role of disentangling within the \ensuremath{\beta}-VAE optimisation objective from the self-supervision aspect of training, we kept as many choices as possible unchanged between the \ensuremath{\beta}-VAE and the AE/VAE baselines: the model architecture, optimiser, learning rate, batch size and number of training steps. The remaining free hyperparameters that could affect the quality of the AE/VAE learnt latent units were the random initialisation seeds, and the number of latent units $N$. The latter was necessary to sweep over explicitly, since AE and VAE models do not have an equivalent to the $\beta$ hyperparameter that affects the intrinsic dimensionality of the learnt representation. Hence, we trained 100 model instances for each of the AE and VAE model classes, with five values of $N$ sampled uniformly without replacement from $N \in [10, 50]$, each trained from twenty random initialisation seed values.

For the Classifier baseline we used the following hyperparameters for the initial selection: five values of $N \in [10, 50]$ sampled uniformly without replacement, as well as a number of learning rate values $\{1e-3, 1e-4, 1e-5, 1e-6, 1e-7\}$ and batch sizes $\{16, 64, 128, 256\}$, resulting in 100 model instances. We trained the models with early stopping to avoid overfitting, and used the classification performance on the validation set to choose the settings for the learning rate and batch size. We found that the values used for training \ensuremath{\beta}-VAE, AE and VAE (learning rate $1e-4$, batch size 16) were also reasonable for training the Classifier, achieving $>95\%$ classification accuracy. Hence, we trained the final set of 450 Classifier model instances with fixed learning rate and batch size, five values of $N \in [10, 50]$ sampled uniformly without replacement and fifty random seeds.

We used FastICA algorithm\cite{hyvarinen2000fastica} to extract ICA units, which is dependent on the random initialisation seed. Hence, we extracted $N \in [10, 50]$ independent components with ten random initialisation seeds each, resulting in 41 ICA model instances. 

The remaining baseline models relied on using a single canonical model instance (VGG and AAM) and/or on a deterministic dimensionality reduction process (PCA, AAM, VGG). Hence, the random seed hyperparameter did not apply to them. In order to make a fairer comparison with the other baselines, we therefore created different model instances by extracting different numbers of representation dimensions with $N \in [10, 50]$, resulting in 41 PCA and VGG (PCA) model instances, and 21 AAM instances (since $N$ needs to split evenly into shape and appearance related units). For the VGG (raw) variant, we first uniformly sampled five values from $N \in [10, 50]$ without replacement to indicate the size of the hidden unit subsets. Then, for each size value we sampled ten random hidden unit subsets without replacement, resulting in 50 VGG (raw) model instances in total.

\paragraph{Model selection based on training performance} For each model class, apart from the deterministic baselines (PCA, AAM and VGG), we selected a subset of model instances based on their training performance. For the \ensuremath{\beta}-VAEs, we used the recently proposed Unsupervised Disentanglement Ranking (UDR) score\cite{Duan_etal_2019} to select 51 model instances with the most disentangled representations (within the top 15\% of UDR scores) for further analysis. For AE baseline, we selected 50 model instances with the lowest reconstruction error per chosen value of $N$. For the VAE baseline we selected 50 model instances with the highest lower bound on the training data distribution per chosen value of $N$. Finally, for the Classifier baseline, we selected 81 models which achieved $>95\%$ classification accuracy on the test set.

\paragraph{Filtering out uninformative models} To ensure that all models used in the final analyses shared at least some information with the recorded neural population, we performed the following filtering procedure. First, we trained Lasso regressors as per Variance Explained section below, to predict the responses of each neuron across the 2,100 faces from the population of latent units extracted from each trained model. We then calculated the mean amount of Variance Explained (VE) averaged across all neurons for each of the models. We filtered out all models where $VE < \overline{VE} -  \text{SD}(VE)$, with $\overline{VE}$ and $\text{SD}(VE)$ represent the mean and standard deviation of VE scores across all models respectively.

The full model selection pipeline resulted in 51 \ensuremath{\beta}-VAE model instances, 50 AE, VAE and ICA model instances, 41 PCA and VGG (PCA) model instances, 22 VGG (raw) model instances, 21 AAM model instances and 64 Classifier model instances that were used for further analyses.

\subsection*{Analysis methods}
\paragraph{Variance explained}
We used Lasso regression to predict the response of each neuron $\mathbf{n}_j$ from model units. We used 10-fold cross validation using standardised units and neural responses to find the sparsest weight matrix that produced mean squared error (MSE) results between the predicted neural responses $\mathbf{\hat{n}}_j$ and the real neural responses $\mathbf{n}_j$ no more than one standard error away from the smallest MSE obtained using 100 lambda values. The learnt weight vectors were used to predict the neural responses from model units on the test set of images. Variance explained (VE) was calculated on the test set according to the following:

\[
\text{VE}_j = 1 - \frac{\sum_i (\mathbf{\hat{n}}_{ij} - \mathbf{n_{ij}})^2 }{\sum_i (\mathbf{n_{ij}} - \mathbf{\overline{n}}_j)^2 }
\]

where $j$ is the neuron index, $i$ is the test image index, and $\mathbf{\overline{n}}_j$ is the mean response magnitude for neuron $j$ across all test images. In order to speed up the Lasso regression calculations, we manually zeroed out the responses of those model units that did not carry much information about the face images. We defined units as ``uninformative'' if their standardised responses had low variance $\sigma^2<0.01$ across the dataset of 2,100 faces. We verified that this did not affect the sparsity of the resulting Lasso regression weights.

\paragraph{Alignment score}
We used the completeness score from\cite{Eastwood_Williams_2018}, referred to as alignment score in text for more intuitive exposition. First we obtained the matrix $R$ necessary for calculating the score by training Lasso regressors to predict the responses of each neuron from the population of model latent units. When calculating completeness against the original neural responses, we followed the same procedure as per the variance explained calculations. When calculating completeness against the artificial (linearly recombined) neural responses, we did not zero out the responses of the ``uninformative'' units, since in this case this procedure affected the sparsity of the resulting Lasso regression weights. Instead, in order to speed up calculations, we reduced the number of cross validation splits from ten to three. The completeness score $C_j$ for neuron $j$ was calculated according to the following:

\begin{equation}
C_j = \rho_j(1 - H(p_j)) 
\end{equation}
\begin{equation}
H(p_j) = -\sum_{d} \ p_{dj} \ log_D \ p_{dj} 
\end{equation}
\begin{equation}
p_{dj} = \frac{R_{dj}}{\sum_d \ R_{dj}}  
\end{equation}
\begin{equation}
\rho_j = \frac{\sum_d R_{dj}}{ \sum_{dj} R_{dj}} 
\end{equation}

where $j$ indexes over neurons, $d$ indexes over model units, and $D$ is the total number of model units. The overall completeness score per model is equal to the sum of all per-neuron completeness scores $C = \sum_j C_j$. See\cite{Eastwood_Williams_2018} for more details.

\paragraph{Unsupervised Disentanglement Ranking (UDR) score}
The UDR score\cite{Duan_etal_2019} measures the quality of disentanglement achieved by trained \ensuremath{\beta}-VAE models by performing pairwise comparisons between the representations learnt by models trained using the same hyperparameter setting but with different seeds. This approach requires no access to labels or neural data. We used the Spearman version of the UDR score described in\cite{Duan_etal_2019}. For each trained \ensuremath{\beta}-VAE model we performed 9 pairwise comparisons with all other models trained with the same $\beta$ value and calculated the corresponding UDR$_{ij}$ score, where $i$ and $j$ index the two \ensuremath{\beta}-VAE models. Each UDR$_{ij}$ score is calculated by computing the similarity matrix $R_{ij}$, where each entry is the Spearman correlation between the responses of individual latent units of the two models. The absolute value of the similarity matrix is then taken $|R_{ij}|$ and the final score for each pair of models is calculated according to:  

\begin{equation}
\label{rel_strength}
\frac{1}{d_a+d_b}\left [  \sum_b  \frac{r_a^2 * I_{KL}(b)}{\sum_a R (a, b)}  +  \sum_a \frac{r_b^2 * I_{KL}(a)}{\sum_b R (a, b)} \right ]  
\end{equation}

where $a$ and $b$ index into the latent units of models $i$ and $j$ respectively, $r_a = \max_a R (a, b)$ and $r_b = \max_b R (a, b)$. $I_{KL}$ indicate the ``informative'' latent units within each model, and $d$ is the number of such latent units. The final score for model $i$ is calculated by taking the median of UDR$_{ij}$ across all $j$.

\paragraph{Average correlation ratio and average unit proportion}
For each neuron we calculated the absolute magnitude of Pearson correlation with each of the ``informative'' model units. We then calculated the ratio between the highest correlation and the sum of all correlations per neuron. The ratio scores were then averaged (mean) across the set of unique model units with the highest ratios, and this formed the \emph{average correlation ratio} score per model. The number of unique model units with the highest ratios divided by the total number of informative model units formed the \emph{average unit proportion} score.

\paragraph{Decoding novel faces from single neurons} We first found the best one-to-one match between single model units and corresponding single neurons. To do this, we calculated a correlation matrix $D_{ij} = \text{Corr}(\mathbf{z}_{i}, \mathbf{r}_{j})$ between the responses of each model unit $\mathbf{z}_i$ and the responses of each neuron $\mathbf{r}_j$ over the subset of 2,038 face images that were seen by both the models and the primates, where $\text{Corr}$ stands for Pearson correlation. We then used Hopcroft-Karp\cite{Hopcroft_Karp_1973} algorithm to find the best one-to-one assignment between each model unit and a unique neuron based on the lowest overall $(1-D_{ij})$ score across all matchings. We used the resulting one-to-one assignments to regress the responses of single latent units from the responses of their corresponding single neurons to the heldout 62 faces, using the same subset of 2,038 face images that were seen by both the models and the primates for estimating the regression parameters. We standardised both model units and neural responses for the regression. The resulting predicted latent unit responses were fed into the pre-trained model decoder to obtain reconstructions of the novel faces. We calculated the cosine distance between the standardised predicted and real latent unit responses for each face (after filtering out the ``uninformative'' units), and presented the mean scores across the 62 held out faces for each model. 

\paragraph{Statistical tests}
We used a two-tailed Welsch's t-test for all pairwise model comparisons.

\clearpage

\begin{figure}[ht]
\begin{center}
\includegraphics[width=1.\textwidth]{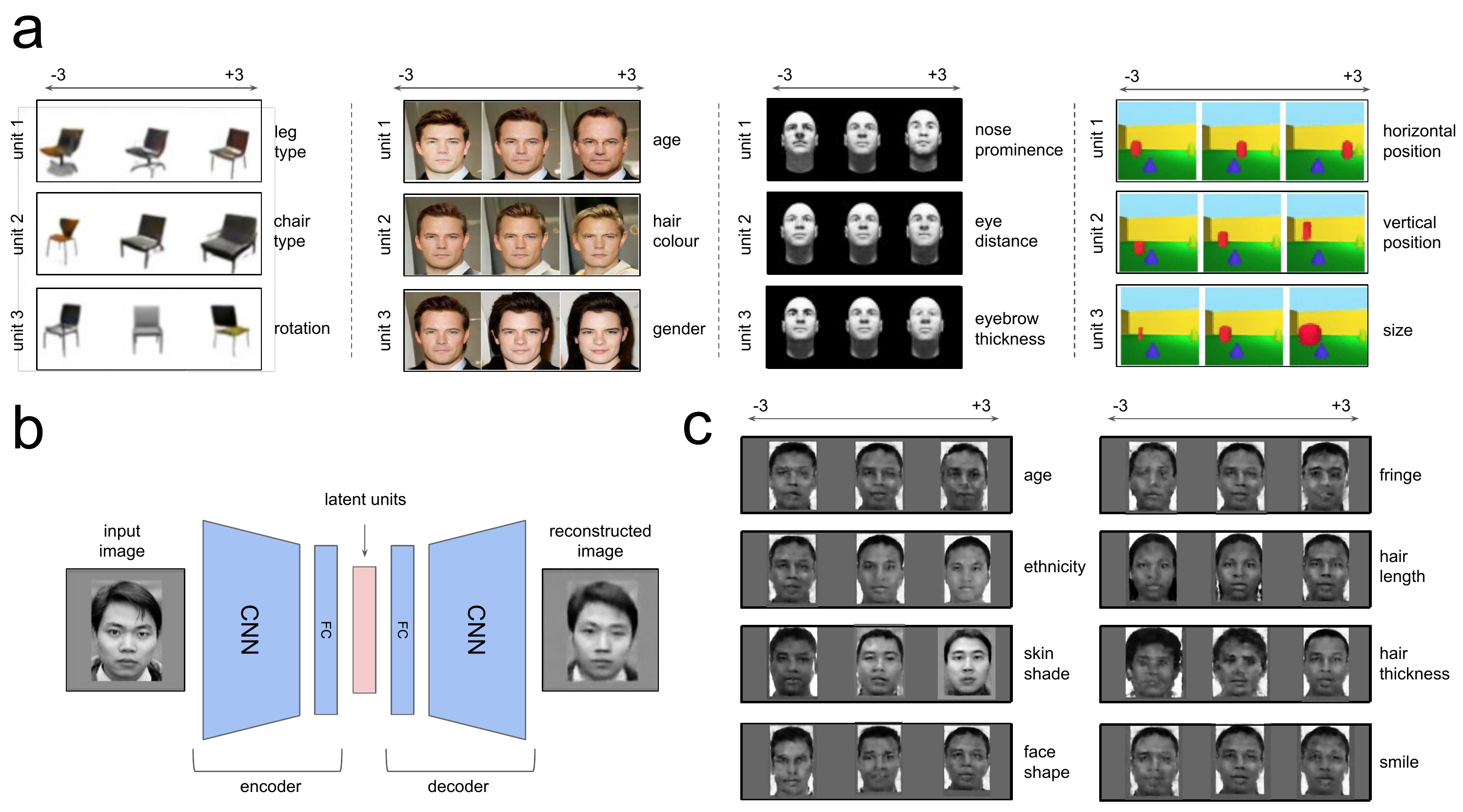}
\caption{\textbf{Disentangled representation learning.} \textbf{a.} Latent traversals used to visualise the semantic meaning encoded by single disentangled latent units of a trained model. In each row the value of a single latent unit is varied between -3 and 3, while the other units are fixed. The resulting effect on the reconstruction is visualised. Each column represents a different model trained to disentangle a different dataset. Aspects of some sub-figures are reproduced with the permission of Burgess et al\cite{burgess2019monet} and Lee et al\cite{lee2020idgan}. \textbf{b.} Schematic representation of a self-supervised deep neural network. The encoder maps the input image into a low dimensional latent representation, which is used by the decoder to reconstruct the original image. Blue indicates trainable neural network units that are free to represent anything. Pink indicates latent representation units that are compared to neurons. CNN, convolutional neural network. FC, fully connected neural network. \textbf{c.} Latent traversals of eight units of a \ensuremath{\beta}-VAE model trained to disentangle 2,100 natural face images. The initial values of all latent units were obtained by encoding the same input image.}
\label{fig1}
\end{center}
\end{figure}

\begin{figure}[ht]
\begin{center}
\includegraphics[width=1.\textwidth]{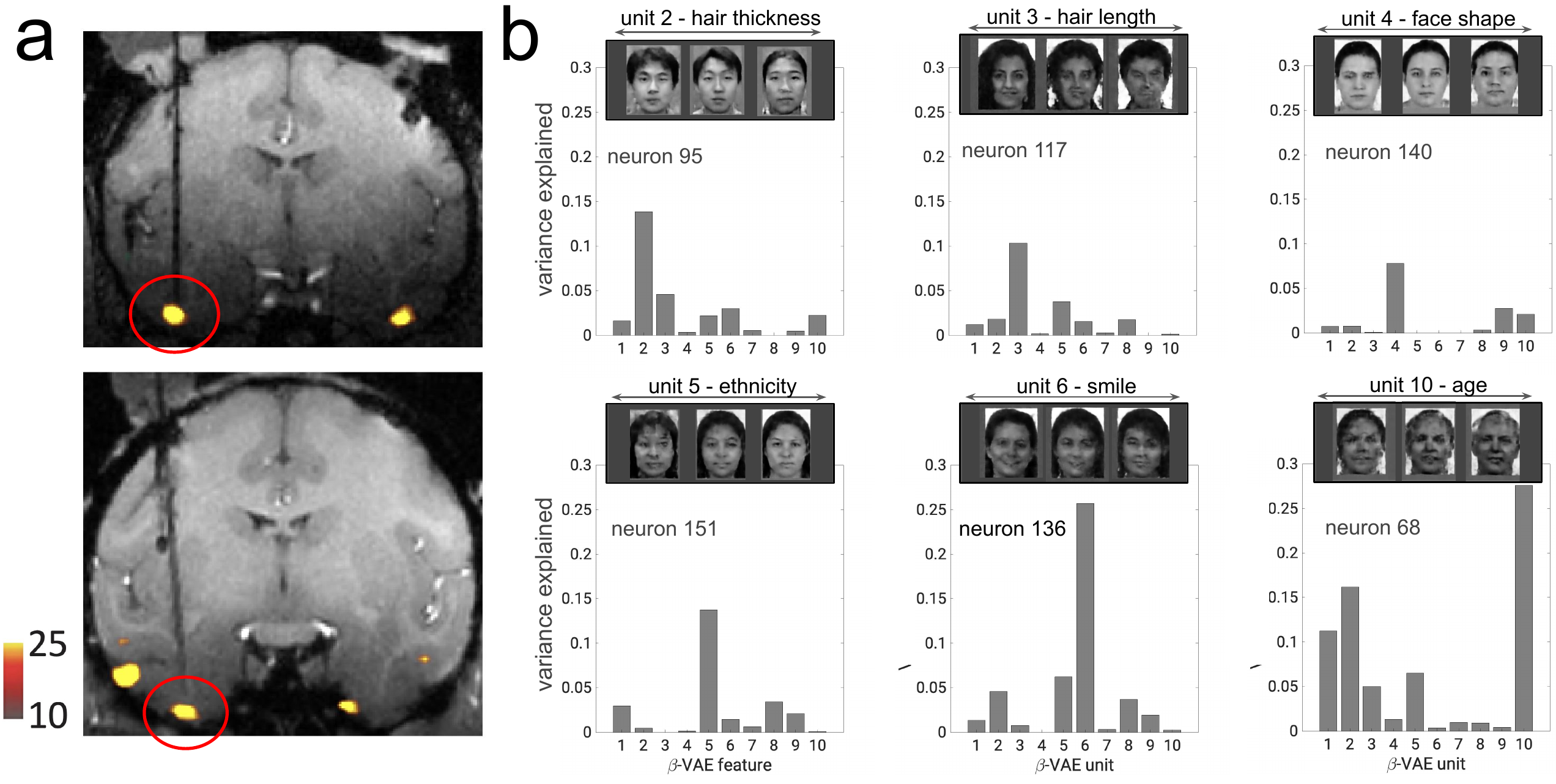}
\caption{\textbf{Responses of single neurons are well explained by single disentangled latent units.} \textbf{a.} Coronal section showing the location of fMRI-identified face patches in two primates, with patch AM circled in red. Dark black lines, electrodes. \textbf{b.} Explained variance of single neuron responses to 2,100 faces. Response variance in single neurons is explained primarily by single disentangled units encoding different semantically meaningful information (insets, latent traversals as in Fig. 1a,c).}
\label{fig2}
\end{center}
\end{figure}

\begin{figure}[ht]
\begin{center}
\includegraphics[width=1.\textwidth]{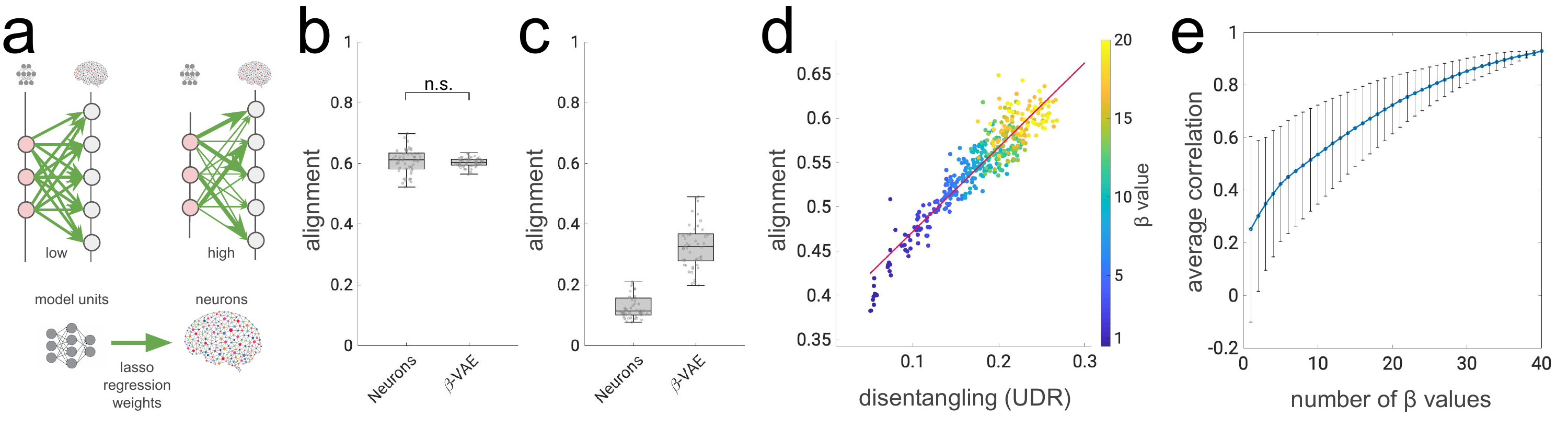}
\caption{\textbf{Strong alignment between single neurons and disentangled units.} \textbf{a.} Schematic of alignment score\cite{Eastwood_Williams_2018,Ridgeway_Mozer_2018}. Green arrows, lasso regression weights obtained from predicting neural responses from model units (thickness indicates weight magnitude). High alignment scores are obtained when per-neuron regression weights have low entropy (one strong weight); high entropy (all incoming weights are of equal magnitude) results in low alignment scores. \textbf{b.} \ensuremath{\beta}-VAE alignment scores match the ceiling provided by subsets of neurons ($p=0.46$, Welsch's t-test). Circles, median alignment per model (n=51) or neuron subsets (n=50). Boxplot center is median, box extends to 25th and 75th percentiles, bars extend to non-outlier data. \textbf{c.} Median alignment scores against artificial neural responses (linear recombination of original neural responses). Boxplot details same as in (\textbf{b}). \textbf{d.} Alignment scores correlate with the disentanglement quality of latent units obtained from 400 \ensuremath{\beta}-VAE models trained with different $\beta$ values (indicated by colour). UDR, Unsupervised Disentanglement Ranking\cite{Duan_etal_2019}, measures the quality of disentanglement, higher is better. Red line, least squares fit (r=0.96, Pearson correlation). \textbf{e.} Running correlation between UDR and alignment scores across subsets of models. Models in each subset were trained with different $\beta$ values, with the number of $\beta$ values in each subset indicated on the x-axis. Rightmost circle, Pearson correlation across 400 \ensuremath{\beta}-VAE models, spanning 40 $\beta$ values as reported in (\textbf{f}). Leftmost circle, average across 40 Pearson correlations, each calculated with 10 models with the a single $\beta$ value. Bars, standard deviation.}
\label{fig3}
\end{center}
\end{figure}

\begin{figure}[ht]
\begin{center}
\includegraphics[width=1.\textwidth]{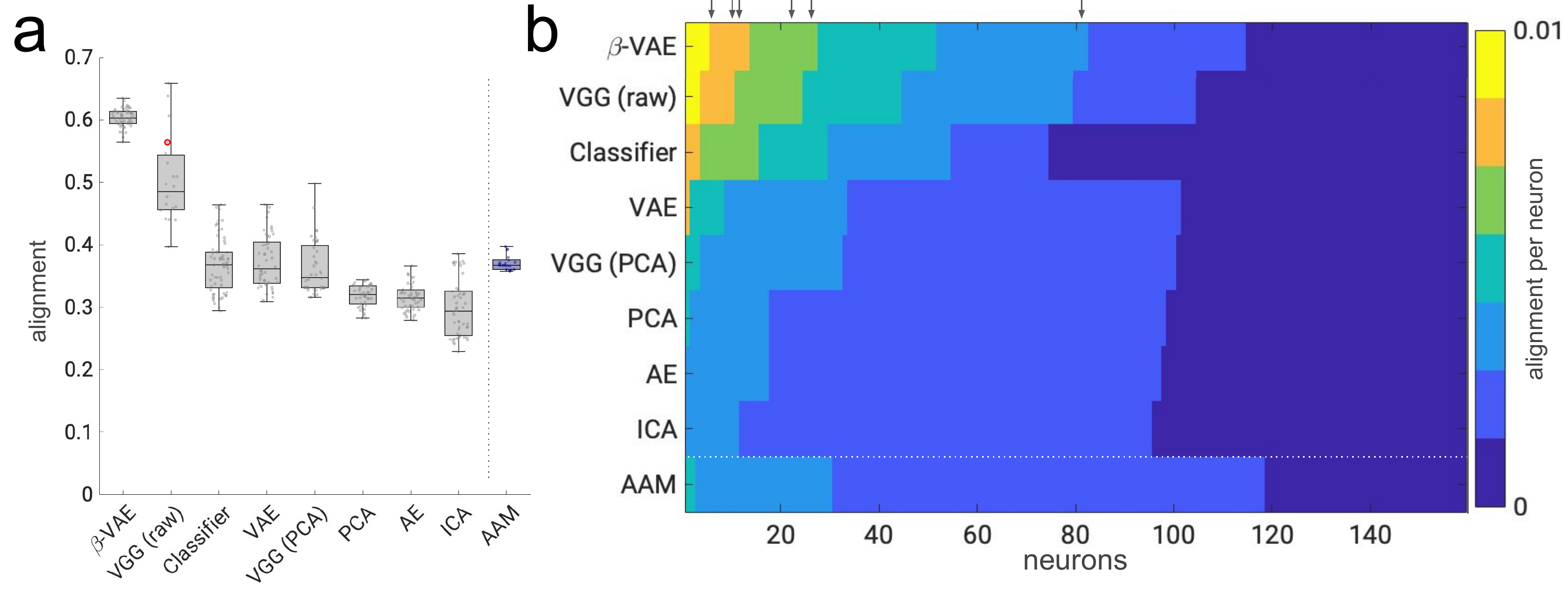}
\caption{\textbf{Disentangled latent units align with single neurons better than baselines.} \textbf{a.} Alignment scores are significantly higher for the \ensuremath{\beta}-VAE than the baseline models and the ``gold standard'' provided by the active appearance model (AAM) (all $p<0.01$, Welsch's t-test). Circles, median alignment per model (\ensuremath{\beta}-VAE, n=51; VGG (raw), n=22; Classifier, n=64; VAE, Variational AutoEncoder\cite{Kingma_Welling_2014}, n=50; AE, AutoEncoder\cite{Hinton_Salakhutdinov_2006}, n=50; VGG (PCA)\cite{Parkhi_etal_2015}, n=41; PCA, n=41; ICA, n=50; AAM, Active Appearance Model\cite{Chang_Tsao_2017}, n=21). Red circle indicates VGG (raw) with all N=4,096 units from the last hidden layer. Boxplot details are the same as in Fig. 3. \textbf{b.} Per neuron alignment scores. Scores are discretised into equally spaced bins. Scores in each row are arranged in descending order. Results from single models, chosen to have the median alignment score. VGG (raw) results are presented from the model that contained all N=4,096 units from the last hidden layer. Arrows point to neurons from Fig. 2b within the \ensuremath{\beta}-VAE alignment scores. }
\label{fig4}
\end{center}
\end{figure}

\begin{figure}[ht]
\begin{center}
\includegraphics[width=1.\textwidth]{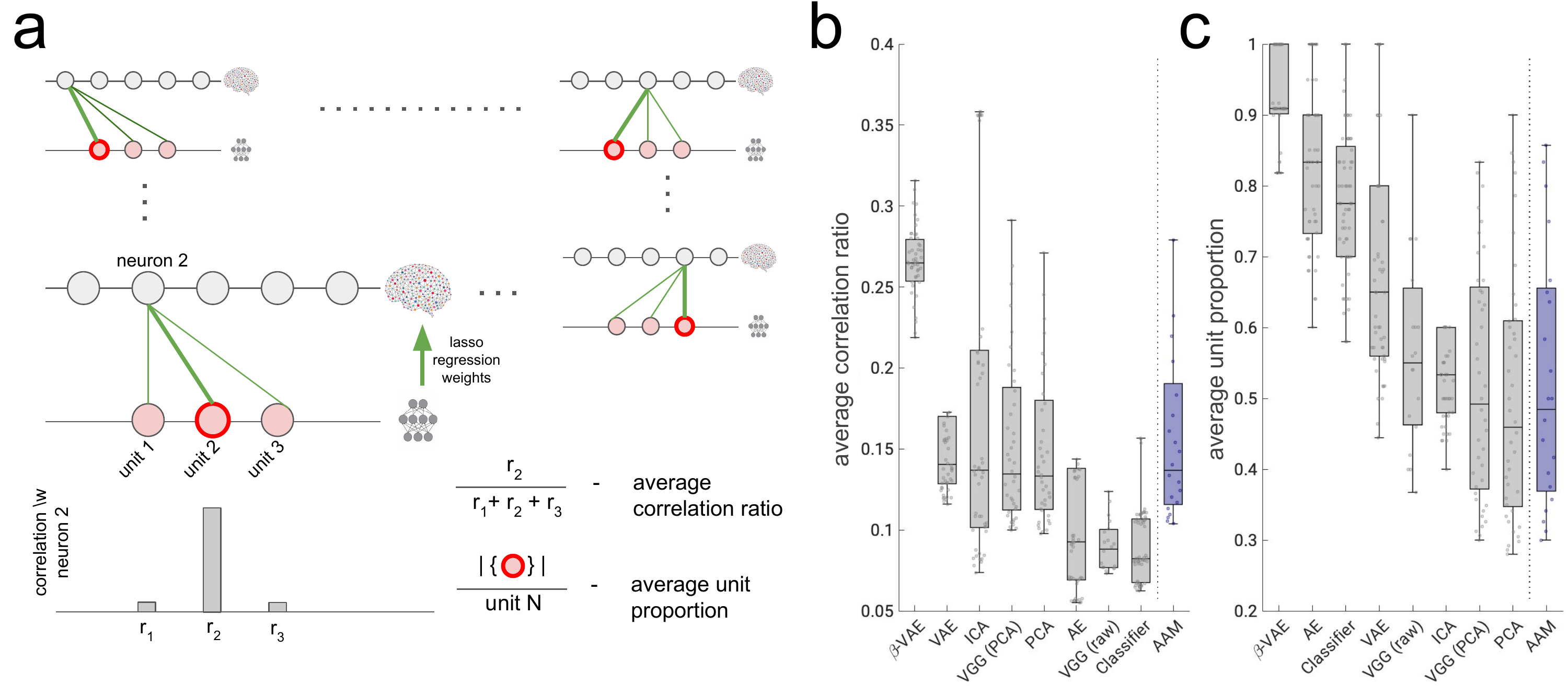}
\caption{\textbf{Disentangled latent units have better diversity and one-to-one correlation with neurons compared to baselines.} \textbf{a.} Schematic of \emph{average correlation ratio} and \emph{average unit proportion} scores. A good computational model for explaining responses of single neurons should allow each neuron (grey circle) to be decodable from a single latent unit (pink circle). Green lines are Lasso regression weights as in Fig. 3a. The response of each neuron should correlate strongly with the response of only one latent unit (grey bars) as measured by the average correlation ratio (higher is better). Different neurons should correlate strongly with diverse single latent units (red circles) as measured by the average unit proportion score (higher is better). \textbf{b.} Average correlation ratio scores are significantly higher for the \ensuremath{\beta}-VAE than the baseline models and the AAM model (all $p<0.01$, Welsch's t-test). Boxplot details are the same as in Fig. 3. \textbf{c.} Average unit proportion scores are significantly higher for the \ensuremath{\beta}-VAE than the baseline models and the AAM model (all $p<0.01$, Welsch's t-test). Boxplot details are the same as in Fig. 3.  }
\label{fig5}
\end{center}
\end{figure}

\begin{figure}[ht]
\begin{center}
\includegraphics[width=1.\textwidth]{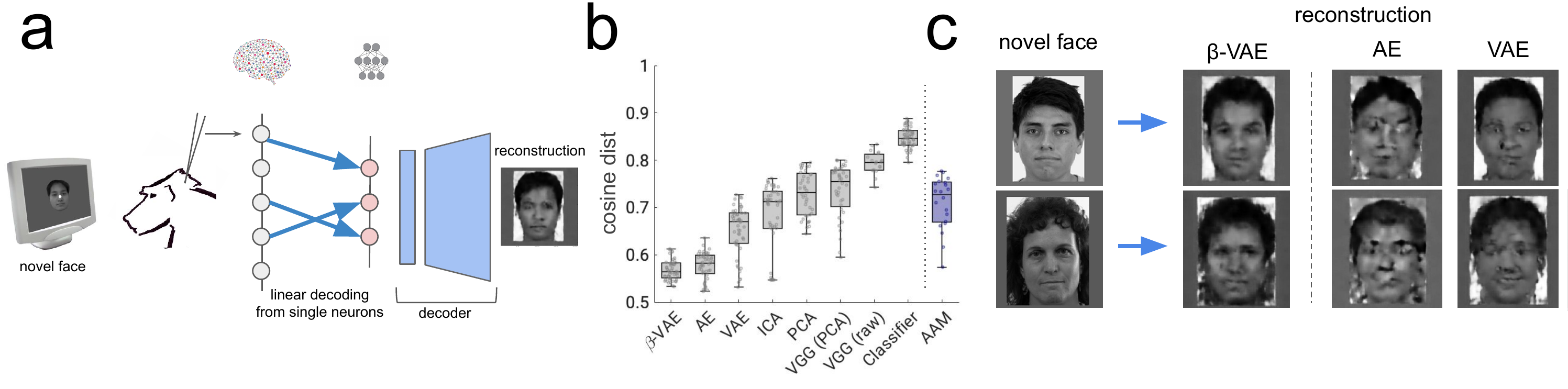}
\caption{\textbf{Reconstructing novel faces from single neurons.} \textbf{a.} Responses of 159 neurons (grey circles) in face patch area AM were recorded while two primates viewed 62 novel faces. One-to-one match was found between each model unit (pink circles) and a corresponding single neuron. Linear regression (blue arrow) was used to decode the responses of each individual model latent unit (pink circles) from the activations of its corresponding single neuron. The pre-trained model decoder was used to reconstruct the novel face. \textbf{b.} Cosine distance between real standardised latent unit responses and those decoded from single neurons are significantly smaller for \ensuremath{\beta}-VAE compared to baseline models and the ``gold standard'' provided by the AAM model (all $p<0.01$, Welsch's t-test). Boxplot details are the same as in Fig. 3. \textbf{c.} \ensuremath{\beta}-VAE can decode and reconstruct novel faces from 12 matching single neurons. The reconstructions are better than those from the closest baselines, AE and VAE, which required 30 and 27 neurons for decoding respectively. The \ensuremath{\beta}-VAE instance was chosen to have the best disentanglement quality as measured by the UDR score; AE and VAE instances were chosen to have the highest reconstruction accuracy on the training dataset.}
\label{fig6}
\end{center}
\end{figure}

\begin{figure}[ht]
\begin{center}
\includegraphics[width=1.\textwidth]{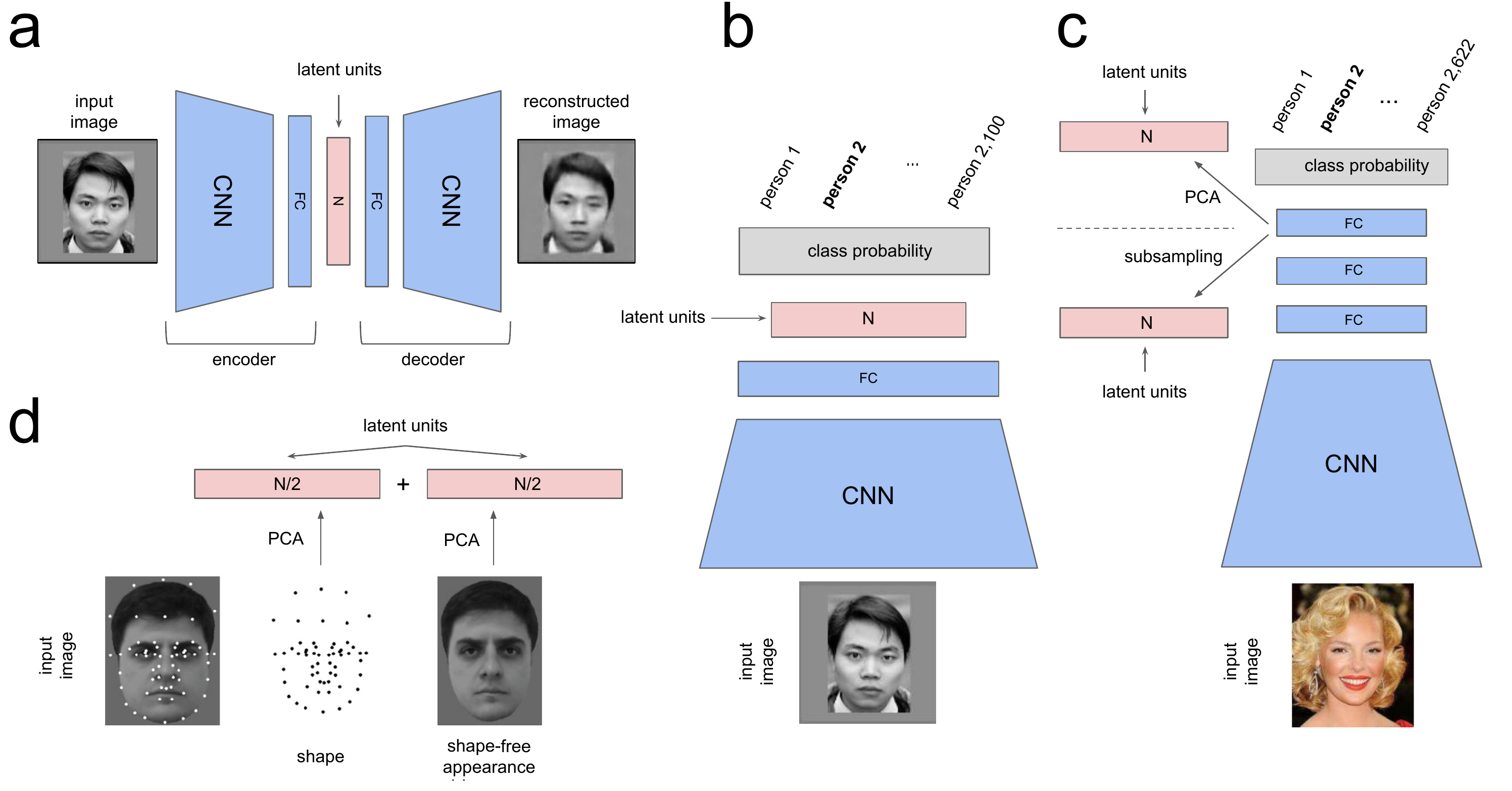}
\caption{\textbf{Schematic of model architectures.} Blue, trainable neural network units free to represent anything. Pink, latent representation units used for comparison with neurons in response to 2,100 face images. Grey, units representing class probabilities. CNN, convolutional neural network. FC, fully connected neural network. N, number of latent units. \textbf{a.} Self-supervised models -- \ensuremath{\beta}-VAE\cite{Higgins_etal_2017}, autoencoder (AE)\cite{Hinton_Salakhutdinov_2006} and variational autoencoder (VAE)\cite{Kingma_Welling_2014,Rezende_etal_2014}. Models were trained on the mirror flipped versions of the 2,100 faces presented to the primates. \textbf{b.} Classifier baseline. Encoder network, same as in (\textbf{a}). Model trained to differentiate between unique 2,100 face identities using mirror flipped versions of the 2,100 faces augmented with 5x5 pixel translations. \textbf{c.} VGG baseline\cite{Parkhi_etal_2015}. Encoder network has larger and deeper CNN and FC modules than in (\textbf{a}) and (\textbf{b}). Representation dimensionality is reduced to match other models either by a projection on the first N principal components (PCs) (VGG (PCA)), or by taking a random subset of N units without replacement (VGG (raw)). VGG was trained to differentiate between 2,622 unique faces using a face dataset\cite{Parkhi_etal_2015} unrelated to the 2,100 faces presented to the primates. \textbf{d.} Active appearance model (AAM)\cite{Chang_Tsao_2017}. Keypoints were manually placed on the 2,100 face images. First N/2 PCs over the keypoint locations formed the ``shape'' latent units. First N/2 PCs over the shape-normalised images formed the ``appearance'' latent units.}
\label{fig7}
\end{center}
\end{figure}

\end{document}